\documentclass[english,journal = nalefd,manuscript = letter, layout = traditional, email = true]{achemso}
\usepackage[T1]{fontenc}
\usepackage[utf8]{inputenc}
\usepackage{amsmath}
\usepackage{amssymb}
\usepackage{graphicx}
\usepackage{esint}

\makeatletter

\title{Electronic transport across realistic grain-boundaries in Graphene}
\author{Fernando Gargiulo}
\affiliation{Institute of Theoretical Physics, École Polytechnique Fédérale de
Lausanne (EPFL), CH-1015 Lausanne, Switzerland}
\email{fernandogargiulo@gmail.com}

\usepackage{babel}
\setkeys{acs}{articletitle = true}
\usepackage[numbers]{natbib}

\makeatother

\usepackage{babel}
\begin{document}
\begin{abstract}
We perform a MonteCarlo simulation in order to study the connection
between the morphology and the transport properties of grain boundaries
(GBs) in graphene. We explore the configurational space of GBs to
generate ensembles of realistic models of disordered interfaces between
graphene misoriented domains. Among other observables, transmission
across GBs has been probed all along the simulation, thus making us
able to establish a connection between averaged transmission and the
topological invariant of GBs, the misorientation angle. We extend
to disordered GBs the remarkable result that the low angle regime
is characterized by a decrease of the individual GB conductance upon
a reduction of the angle, as first found for periodic GBs \cite{gargiulo_topological_2013}.
However, we explored a comprehensive range of misorientation angles
such that our results should serve as a starting data set to study
the effect of polycrystallicity on transport in large samples. 
\end{abstract}
\maketitle

\section*{Introduction}

Graphene is the first genuine two-dimensional material ever unearthed,
being a one-atom thick sheet of graphite \cite{novoselov_electric_2004,geim_rise_2007}.
As a consequence, its novelty, soon supported by the fascinating physical
properties which it has been showing \cite{castro_neto_electronic_2009},
has rapidly appealed many researchers. After almost a decade of intensive
studies one can hardly find a field of condensed matter Physics which
has not been touched with reference to graphene. The effect of polycrystallicity
on graphene physical properties doesn't make exception. Indeed, it
is now established that grain-boundaries - the topological defects
characteristic of polycrystalline materials - are ubiquitous in extended
graphene samples grown, for example, by chemical vapor deposition
\cite{coraux_structural_2008,yazyev_topological_2010,an_domain_2011,huang_grains_2011,kim_grain_2011}.
Their strong impact over electronic, thermal and mechanical properties
is nowadays out of debate \cite{tapaszto_mapping_2012-1,rasool_measurement_2013,lee_high-strength_2013}.
However, for electronic transport in polycrystalline graphene is far
from being achieved and, in particular, theoretical estimates of the
intrinsic electronic performances of realistic polycrystalline samples
are lacking \cite{das_sarma_electronic_2011,peres_colloquium:_2010,radchenko_effect_2013,tsen_tailoring_2012,vancso_electronic_????}.
This is surprising since the technological applications for which
graphene is expected to be a promising candidate (electronics, clean
energy and related domains) require large scale industrial processes
- e.g. chemical vapor deposition technique - which mostly end up with
the realization of polycrystalline graphene samples, as confirmed
by numerous recent experiments \cite{yu_control_2011,banhart_structural_2011}.

The first theoretical studies on polycrystalline graphene have regarded
grain boundaries as arrays of dislocations, that is, adopting the
theory of Read and Shockeley \cite{read_dislocation_1950,yazyev_topological_2010}.
Both dislocations and grain-boundaries, differently from point defects,
are topological meaning that no local modification of the atoms network
can eliminate them. This distinctive nature of these defects, combined
with the characteristics of pristine graphene, is at the origin of
rich as well as non trivial transport features . A work focusing on
periodic structures have unveiled the existence of a class of grain-boundaries
for which a full suppression of the low-energy conductance occurs,
consequence of momentum conservation \cite{yazyev_electronic_2010}.

A recent study has explored the more general situation of those periodic
grain-boundaries in which no symmetry-related selection applies, also
addressing the effect of perturbations to the periodic order \cite{gargiulo_topological_2013}.
This has provided a general picture of transport in low angle grain-boundaries,
that is, the ones which allow the fewest arrangements of the defective
rings and, consequently, do not bear a high degree of disorder \cite{coraux_structural_2008,yazyev_topological_2010}.
On the other hand, large angles grain-boundaries are often highly
disordered, that is, many defects arrangements are compatible with
a given misorientation angle \cite{huang_grains_2011,kim_grain_2011,banhart_structural_2011}.
This poses a serious obstacle to an understanding of the effect of
a single grain-boundary on charge transport based on simple and general
arguments.

In this work, we address the problem of estimating the conductance
of disordered grain-boundaries numerically, this approach being the
only possible. Having a collection of realistic grain-boundaries for
a given misorientation angle is propaedeutic to the calculation of
any physical observable. We have chosen to employ a MonteCarlo simulation
in order to explore the configuration space of the interface between
two misaligned domains. This has provided bunches of defective structures
not selected randomly but weighted by their formation energy with
respect to the corresponding ordered low-energy configuration . Thereafter,
we have sampled several observables to have a complete picture of
both the morphology (number of atoms of the rings, atomic connectivity,
formation energy), and the spectral and transport properties (DOS,
Transmission, Conductance). The statistical analysis of the data has
put the connection of the average conductance of a single grain-boundary
and its misorientation angle on a quantitative basis. 

As an important ingredient to the transport theory, it has been recently
showed that the presence of multiple grain-boundaries, which is expected
in polycristals, leads to a simple law of direct proportionality between
the conductance of the sample and the average linear size of the single
grain \cite{van_tuan_scaling_2013}. This can be easily interpreted
as the emergence of ohmic behavior induced by the presence of multiple
grain-boundaries.  At this point, our quantitative estimates for
the conductance across a single grain-boundary combined with the knowledge
of the transport regime let us glimpse the opportunity for a multiscale
determination of the intrinsic transport performances in large area
polycrystalline samples. A very minimal, although meaningful, illustrative
example is given at the end of this work.

\section*{Description of the work}

\begin{figure}
\includegraphics[width=12cm]{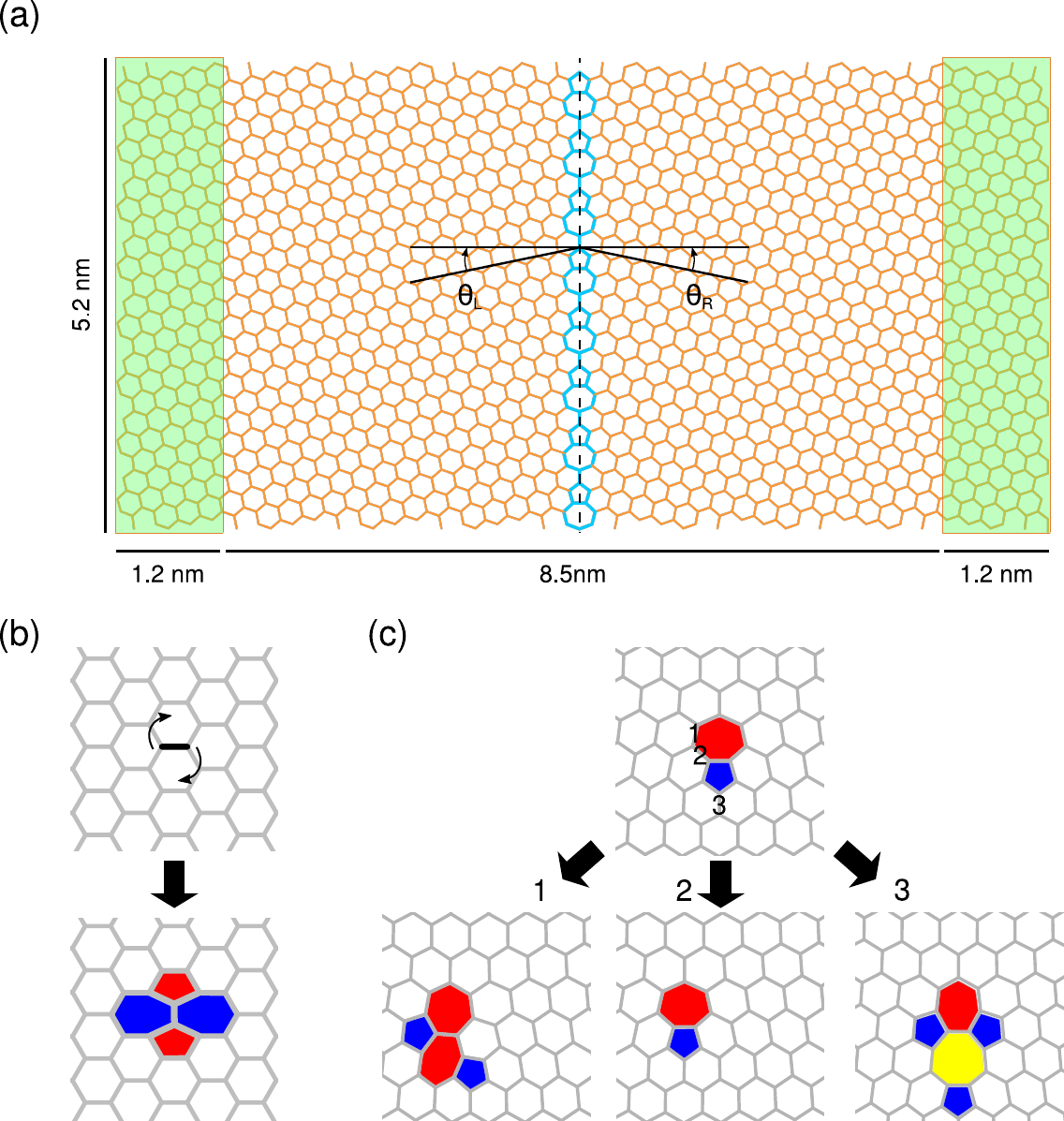}\caption{\label{fig:Model}Sketch of the simulation setup and MC move outcomes.
Panel (a) presents the initial configuration of a two-domains system
with a misorientation angle $\theta=\theta_{\textrm{L}}+\theta_{\textrm{R}}=21.2^{\circ}$.
Both the directions parallel and orthogonal to the GB are periodic.
The atoms in the green regions are kept fixed along the simulation.
Panel (c) shows the effect of a bond rotation in a pristine graphene
area which results in the formation of a Stone-Wales defect. Panel
(d) illustrates three relevant outcomes of the bond rotation around
a 5-7 pair: pair cration (1), glide (2), a higher energy formation
including an eight member ring (3).}
\end{figure}

In order to perform a MonteCarlo simulation two ingredients are necessary:
the basic move and the acceptance criteria. We have chosen the Wooten-Winer-Waeire
move \cite{wooten_computer_1985}consisting in the rotation of two
bound atoms by $90^{\circ}$ as illustrated in Fig.\ref{fig:Model}b.
The system is therefore relaxed by minimizing a classical potential
suited for carbon \cite{los_intrinsic_2003}. Within a pristine area
of graphene this move results in the creation of the so-called Stone-Wales
defect (see Fig. \ref{fig:Model}b), characterized by a high formation
energy of about $5\,\mathrm{\textrm{eV}}$ \cite{li_defect_2005}.
When the rotation is done in the vicinity of a dislocation (i.e. a
pentagon-heptagon pair) it often represents an energetically low-cost
move allowing for the evolution of the defect (see Fig.\ref{fig:Model}c).
In the proximity of a grain-boundary - we remark that a GB can be
thought as an array of dislocations - the move can even lead to a
structure with a lower energy, thus disclosing the exploration of
the configuration space. Therefore, our final choice is to rotate
bonds which connect at least one atoms owning to the GB. We adopt
the Metropolis scheme as the acceptance criteria \cite{metropolis_equation_1953}.
It is important to say that, for our purpose, the MC simulation is
not intended as a tool to obtain a thermodynamic ensemble of configurations.
Indeed, the growth of the grains in a chemical vapor depositions,
leading to the formation of boundaries, happens in conditions which
are out of thermodynamic equilibrium. In our case, the simulation
rather consists in a tool to collect grain-boundary configurations....
In this spirit, it must not confuse that the main temperature chosen
for the simulation $T=5000K$ is close to the melting temperature
for graphene \cite{zakharchenko_melting_2011}. In fact, our simulation
involves only few degrees of freedom in the boundary region and the
energy corresponding to the temperature ($=0.43\,\textrm{eV}$) divided
by the average distance between two carbon atoms ($1.42\,\textrm{\AA}$)
is comparable to the typical formation energy of a GB ($0.2-0.8\,\textrm{eV/\textrm{\AA}}$).
In other words, we have chosen a temperature such that the system
has a significant probability to assume distinct configurations along
the simulation. After each move, the system is relaxed by mean of
a classical force field. The minimized energy is then employed in
the Metropolis scheme. The coherent conductance across the GB is numerically
assessed by mean of the Landauer-B\textbackslash{}''\{u\}ttiker theory,
in which the conductance $G(E)$ at a given energy $E$ is proportional
to the transmission $T(E)$ as $G(E)=G_{0}T(E)$, with $G_{0}=2e^{2}/h$
being the conductance quantum. We use a two-terminal device configuration
in which contacts are represented by semi-infinite ideal graphene
leads. More details can be found in the Methods section. 

\section*{Results and discussion}

\begin{figure}
\includegraphics[width=14cm]{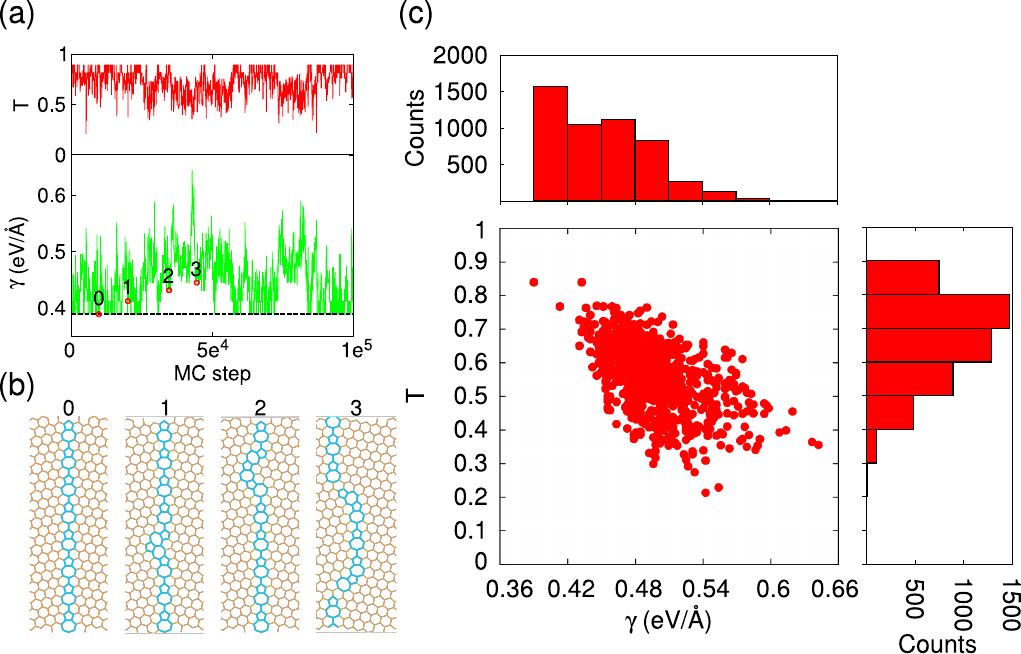}\caption{\label{fig:Evolution}Observables sampled during the simulation of
the $\theta=21.2^{\circ}$ GB. (a) Evolution of the formation energy
and the conductance integrated between $-0.5\,\textrm{eV}$ and $0.5\,\textrm{eV}$.
(b) Ground state and three low-energy configurations extracted from
the simulation. The corresponding energies are highlighted in the
upper plot by red circles. (c) The main frame contains a scatter plot
of conductance $vs.$ formation energy. All conductances are normalized
by the value pertaining to a pristine graphene sheet of the same size
of the polycrystalline sample.}
\end{figure}

We have performed simulations for starting from 8 symmetric grain-boundaries
covering a range of misorientation angle $\theta\in\left[7.3{}^{\circ},51.5^{\circ}\right]$
and 1 asymmetric with $\theta=30^{\circ}$. Data collected along a
typical simulation are presented in Fig.\ref{fig:Evolution}. The
starting configuration is the left one sketched in panel b. It consists
in a $\theta=21.2^{\circ}$ GB made of $8$ pentagon-heptagon pairs. 

The evolution of the formation energy, plotted in \ref{fig:Evolution}a
shows that there are several low-energy configuration occurring frequently.
Three of these disordered GBs are sketched in \ref{fig:Evolution}b,1-3.
From a comparison with the initial ordered configuration (\ref{fig:Evolution}b,0)
it can be seen that each of them results from a combination of glides,
creations or annihilations of pentagon-heptagon pairs. These transformations
have been recognized as the ones responsible for the life cycle of
dislocations and for the evolution of grain-boundaries\cite{lehtinen_atomic_2013,kurasch_atom-by-atom_2012}.
An histogram showing the distribution of the rings is contained in
figure S2. The absence of members with less than 5 members and the
rare occurrence of 8-membered rings has to be put in connection with
the high energy cost of those defects (Refs.??). As a general fact,
in all our simulations, the GBs lying in the low energy region, are
formed by an equal number of 5 membered and 7-membered rings. All
along the evolution of the GBs the coordination number of the atoms
is identically equal to 3 excluding few very high-energy (and rare)
configurations in which four-coordinated or two-coordinated carbon
atoms may appear. A mismatch between the number of pentagons and heptagons
of a GB also implies a higher formation energy. However, even in the
high energy region, the typical shape of a grain boundary tends to
be meandering but still continuous. This is in accordance with what
was found in experimental atomic resolution imaging of GBs \cite{huang_grains_2011,kim_grain_2011}.
Together with the energy, \ref{fig:Evolution}a shows the evolution
of the conductance in an energy window $1\,\textrm{eV}$ wide centered
around the Fermi level of pristine graphene. For sake of clearness
the conductance $G$ has been normalized by the conductance $G_{\mathrm{P}}$
of a pristine graphene sample of the same size of the sample employed
in the simulation.  \ref{fig:Evolution}c contains a statistical
analysis of the data. A broad distribution characterized the histogram
of the conductance, although it still shows the persistence of high
transmitting configurations. The main panel clearly indicates the
existence of some inverse correlation between the conductance and
the energy, meaning that, on average, a higher energy corresponds
to a lower conductance. This can be explained by appreciating the
fact that configurations with high energy correspond to more disordered
GBs constituted by a larger number of non six-membered rings, that
is, by a larger number of scatterers for the charge carriers.

\begin{figure}
\includegraphics[width=8.2cm]{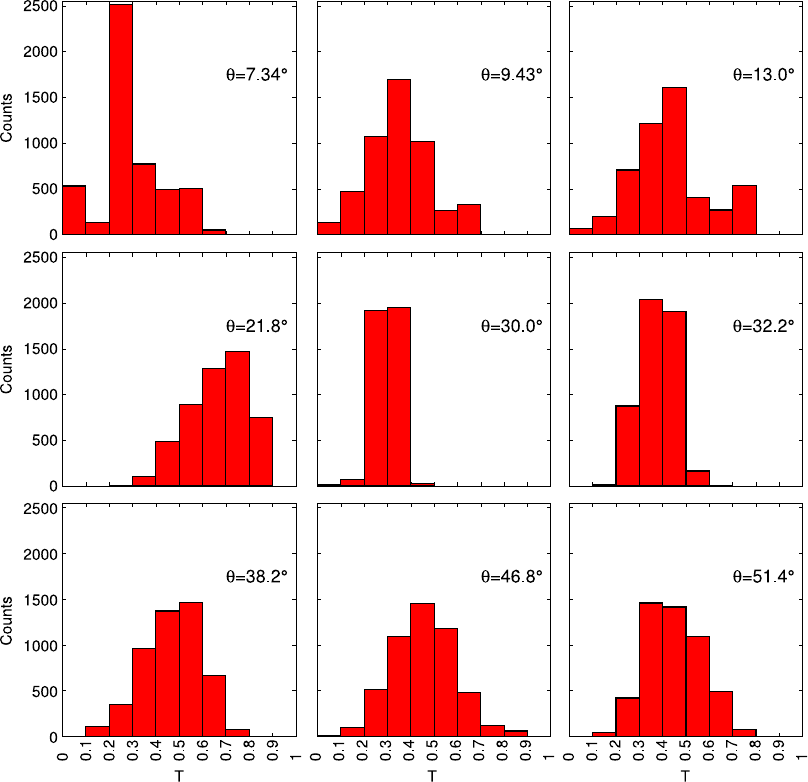}\caption{\label{fig:Histograms}Distributions of the conductance for different
systems characterized by the misorientation angle $\theta,$ as reported
in the panels. The unit adopted for the conductance is the same as
Fig.~\ref{fig:Evolution} }
\end{figure}

Fig.\ref{fig:Histograms} reports the distribution of the integrated
conductance for the different GBs. Going through increasing misorientation
angles, one sees that the distributions evolve almost continuously
with a sudden change registered between $21.8^{\circ}$ and $30^{\circ}$,
giving an indication for an abrupt mutation of the conductance trend.
This strong suppression of the conductance cannot be attributed to
the asymmetry of the $\theta=30^{\circ}$ GB. In fact, the GB with
the next larger angle $\theta=32.2^{\circ}$ has a similar distribution,
though being symmetric i.e. presenting a radically different arrangement
of the defects. This has to be considered an indication for the misorientation
angle $\theta$ to be the main variable which determines the transport
across the GB.

When conductance is evaluated at energies close to the Fermi level
($E=0.02\,\textrm{eV}$) the distributions are broader, although the
average conductance is slightly higher. This instability at the Fermi
level is reduced when the conductance is integrated over an energy
interval. The existence of correlations between formation energy and
conductance is also more evident after the integration. 

\begin{figure}
\includegraphics[width=8.2cm]{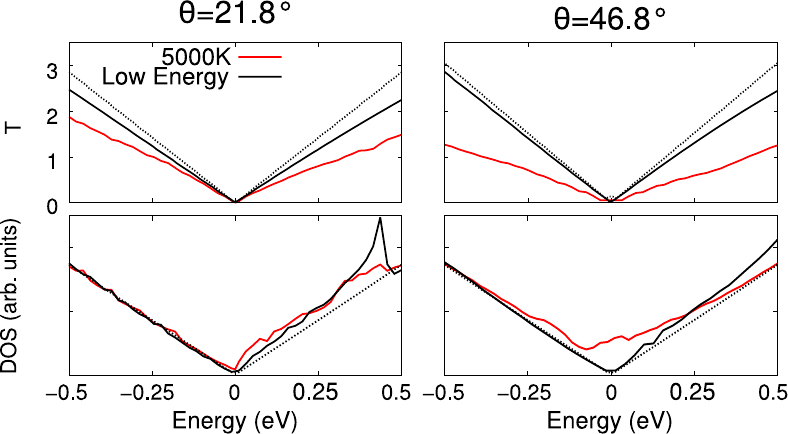}\caption{\label{fig:T_DOS}Average transmission (upper panels) and density
of states (lower panels) as a function of Energy for two representative
systems characterized by $\theta=21.8^{\circ}$and $\theta=46.8^{\circ}.$
The dashed lines represent the calculated values for a pristine graphene
sheet of the same size, the red curves are the simulation averages
of the observables and the black solid lines represent the calculated
values for the structural ground state configuration.}
\end{figure}

A more detailed picture of the effect of the presence of disorder
is obtained by looking at transmission and density of states (DOS)
as a function of energy, reported in Fig.~\ref{fig:T_DOS}. Independently
of the energy, the average transmission is significantly reduced with
respect to that of the ground (ordered) state, that is, the effect
of disorder in GBs is to add further charge carrier backscattering
with respect to the ordered case. This feature is common to the majority
of the systems, the only exception being the case in which the ordered
configuration has already a low conductance compared to pristine graphene.
In such a situation the average conductance doesn't differ significantly
(see S3). 

\begin{figure}
\includegraphics[width=8.2cm]{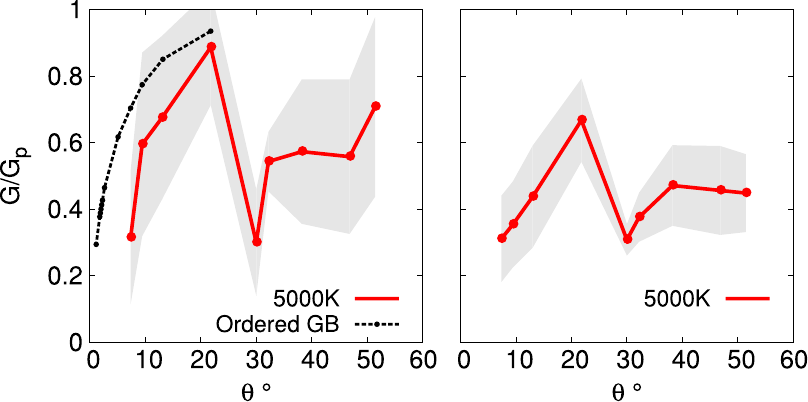}\caption{\label{fig:T_angle}Conductance as a function of the misorientation
angle $\theta$. In panel (a) the conductance is calculated at low
energy $E=2.7\,\textrm{meV}$, whereas in panel (b) it is integrated
between $-0.5\,\textrm{eV}$ and $0.5\,\textrm{eV}$. The red points
are the simulation averages, the width of the gray area corresponds
to the standard deviation of the distributions, the black dashed lines
represent the conductance for the ground state configurations in the
small angle regime. }
\end{figure}

The effect of a finite simulation temperature on the averaged DOS
can be summarized in two aspects. First, characteristic peaks of the
ground state configuration get smeared out. Secondly, there is an
increase of the spectral weight around $E=0$,  

We stress again that the conductance of disordered GBs has a clear
although, not trivial, dependence on the misorientation angle $\theta$.
This dependence has been put in evidence in Fig. \ref{fig:T_angle}
for two situation. In the left panel the average conductance at low
energy ($E=0.02\,\textrm{eV}$) is plotted as a function of the angle
$\theta$. In order to make a comparison possible with the case of
ordered GBs, we have added an analogous curve for the ground state
conductance in the low-angle region. This latter case is characterized
by a suppression of the conductance upon a reduction of the misorientation
angle \cite{gargiulo_topological_2013,mesaros_electronic_2010}. This
counter intuitive behavior has been explained from the point of view
of resonant backscattering induced by quasi-localized states that
get closer to the Fermi energy upon a reduction of $\theta$. It becomes
immediately evident that this trend is inherited by the average conductance
of disordered GBs, showing a maximum at $\theta=21.8^{\circ}$. The
region $30^{\circ}<\theta<60^{\circ}$ , on the other hand exhibits
a less clear trend. In any case, conductance is affected by large
relative fluctuations of the order of $20-50\%$ . This instability
at low energy is sensibly improved by integrating the conductance
over the usual range $[-0.5\,\textrm{eV},0.5\,\textrm{eV}]$(\ref{fig:T_angle}b).
Although after integration the values of average conductance are lower
for most of the angles, fluctuations get reduced by approximately
a factor $2$ allowing to appreciate two well separated trends. Again,
starting from the maximum achieved for $\Theta=21.8^{\circ}$ and
going toward low angles, the conductance decreases reproducing, initially,
the behavior of the ordered GBs. Nevertheless, for a value of $\theta$
around $10^{\circ}$, the conductance of the ordered GB suddenly rise
and is supposed to approach $1$ in the limit of $\theta\rightarrow0$,
whereas the averaged conductance follows an almost straight line until
$\theta=7.54^{\circ}$. Although computational limitations prevents
us from reducing the angle further, we expect the average conductance
to approach the conductance of the ordered GBs in the limit of zero
angle. In this limit, indeed, the ordered low energy configuration
has to be predominant since it is constituted by largely separated
pentagon-heptagon pairs and any modification has a high energy cost.
As a consequence the average is dominated by the contribution of the
low energy configuration.

A different trend characterizes the region $30^{\circ}<\theta<60^{\circ}$.
After a minimum for $\theta=30^{\circ}$, the conductance increases
before stabilizing around $G\simeq0.45G_{p}$. ... . Overall, the
effect of disordered GBs is to reduce the conductance of pristine
graphene to about the $40\%$ of the conductance of pristine graphene.

\begin{figure}
\includegraphics{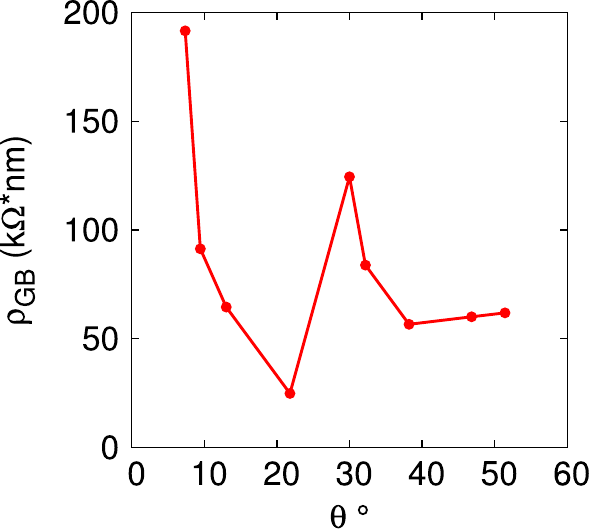}

\caption{\label{fig:Rho_GB}Width independent resistance $\rho_{\mathrm{GB}}$
averaged between $-0.5\,\textrm{eV}$ and $0.5\,\textrm{eV}$ as a
function of the misorientation angle $\theta$.}
\end{figure}

Based on the knowledge of the conductance (or, equivalently, resistance)
of a single grain boundary one can address the effect on electronic
transport of a distribution of grain boundaries typical of a polycrystalline
sample. As showed in ref.\cite{van_tuan_scaling_2013}, the effects
of many individual grain boundaries add-up incoherently meaning that
quantum interference effects are negligible and the resistance due
to individual GBs is additive. The resistance of a two-terminal configuration
$R=G^{-1}$ can be viewed as arising from two contributions\cite{datta_electronic_1997}:
\begin{equation}
R=R_{\mathrm{P}}+R_{\mathrm{GB}}=G_{\mathrm{P}}^{-1}+G_{\mathrm{P}}^{-1}\left(\frac{G_{\mathrm{p}}-G}{G}\right)
\end{equation}
where $R_{\mathrm{P}}$ is the resistance due to the semi-infinite
graphene contacts (i.e. the resistance of a pristine sample) and $R_{\mathrm{GB}}$
is the resistance of the scattering source - the GB in our case -
with the property of being additive. Since the resistance is inversely
proportional to the transverse width $W$, it is convenient to introduce
a width-independent grain boundary resistance $\rho_{\mathrm{GB}}=R_{\mathrm{GB}}*W$
\cite{tsen_tailoring_2012}. Our results recast in terms of $\rho_{\mathrm{GB}}$
(see Fig.~\ref{fig:Rho_GB}) show an even stronger dependence on
the misorientation angle. However, the values of $\rho_{\mathrm{GB}}$
presented in Fig.~\ref{fig:Rho_GB} should be considered as lower
boundary estimates since they not account for potential barriers induced
by grain-boundaries \cite{tapaszto_mapping_2012,koepke_atomic-scale_2013,ihnatsenka_electron_2013,clark_spatially_2013}
, incoherent processes triggered by the presence of grain boundaries
or ... . 

Finally, if one accounts for an average value $\left\langle \rho_{\mathrm{GB}}\right\rangle \simeq80k\Omega*nm$,
a rough estimate for the intrinsic resistance of a polycrystalline
sample of width $W$ and length $L$ with a linear density of grain
boundaries $n$ is given by 
\begin{equation}
R=\frac{nL}{W}\left\langle \rho_{GB}\right\rangle 
\end{equation}
This formula neglects all details of the angle distribution in realistic
GBs, nevertheless, it provides an order of magnitude for the contribution
of coherent backscattering due to GBs in polycrystalline graphene.
For a square sample with a density $n=0.1nm^{-1}$(?) , this contribution
amounts to $\rho_{\mathrm{GB}}\simeq$8k$\Omega$ .

In conclusion, we have addressed the issue of electronic transport
across disordered grain-boundaries combining a MonteCarlo simulation
for the grain boundary configurations and transport calculations based
on Landauer-B\:{u}ttiker theory. The disordered averaged conductance
exhibits a clear dependence on the misorientation angle which is insensitive
to the disorder. The low-angle regime reproduces the reduction of
the conductance already seen in periodic GBs \cite{gargiulo_topological_2013},
the minimum of conductance is achieved for $\theta=30^{\circ}$ whereas
a saturation characterizes the region $\theta\in\left[30^{\circ},60^{\circ}\right]$.
Based on these results, an estimation of the contribution to total
resistance ascribable to the presence of grain boundaries is formulated.

\section*{Methods}

The starting structural models are rectangular with the GB parallel
to a side whose length is $\simeq6\,\textrm{nm}$, whereas the perpendicular
side measures $\simeq10\,\textrm{nm}$. The total number of atoms
lies between $1848$ and $2144$ atoms (all starting configurations
can be found in Fig. S1). Initially, all systems are relaxed in both
the atomic and the cell degrees of freedom. After each move the structure
is relaxed keeping the cell parameters fixed. Relaxations are achieved
by minimizing the structural energy calculated by means of the classical
potential LCBOP\cite{los_intrinsic_2003}, as implemented in the Open
Source code LAMMPS \cite{_lammps_????,plimpton_fast_1995}. The LCBOP
potential has been selected among available alternatives after the
comparison of the formation energies of several defects (both local
and topological) with DFT results. For the description of low energy
charge carriers we adopt a next-neighbor Tight-Binding Hamiltonian
restricted to $\pi$ orbitals in which the hopping term $t$ is set
to $2.7\textrm{\,\ eV}$ and the energy reference is fixed in such
a way that the on-site energy $\varepsilon_{0}$ vanishes: $H=-t\sum_{\left\langle i,j\right\rangle }a_{i}^{\dagger}a_{j}+\mathrm{H.c.}$
where $a_{i}^{\dagger}$($a_{i}$) is the creation (annihilation)
operator of an electron at the site $i$. Transmission is evaluated
by mean of the Non-Equilibrium Green's function $G_{\mathrm{S}}$
of the scattering region containing the GB. The transmission is a
function of the transverse momentum $k_{\Vert}$ (defined for the
supercell) and the energy $E$: $T(k_{\Vert},E)=\mathrm{Tr}\left[\Gamma_{\mathrm{L}}(k_{\Vert},E)G_{\mathrm{S}}^{\dagger}(k_{\Vert},E)\Gamma_{\mathrm{R}}(k_{\Vert},E)G_{\mathrm{S}}(k_{\Vert},E)\right]$.
$G_{\mathrm{S}}$ is defined as $G_{\mathrm{S}}(k_{\Vert},E)=\left[E^{+}I-H_{\mathrm{S}}-\Sigma_{\mathrm{L}}(k_{\Vert},E)-\Sigma_{\mathrm{R}}(k_{\Vert},E)\right]^{-1}$,
the coupling matrices $\Gamma$ for the left and right lead are given
by $\Gamma_{\mathrm{L}\left(\mathrm{R}\right)}(k_{\Vert},E)=i\left[\Sigma_{\mathrm{L}\left(\mathrm{R}\right)}(k_{\Vert},E)-\Sigma_{\mathrm{L}\left(\mathrm{R}\right)}^{\dagger}(k_{\Vert},E)\right]$,
$H_{\textrm{S}}$ is the Hamiltonian of the scattering region, $\Sigma_{\mathrm{L(R)}}$
are the self-energies coupling the scattering region to the leads
and $E^{+}=E+i\eta I\;(\eta\rightarrow0^{+})$. Conductance $G$ is
obtained from transmission as $G=G_{0}\int_{V_{\textrm{L}}}^{V_{\textrm{R}}}dE\int_{1BZ}T\left(k_{\parallel},E\right)$.
For the integrals, $11$ independent $k$-points and $48$ energy-points
have been used, respectively. DOS has been calculated as $DOS\left(E\right)=-\frac{1}{\pi}\int_{1BZ}\textrm{Im}(G_{\textrm{S}}\left(E^{+},k_{\parallel}\right))$
and the integral has been discretized over a $21$ k-points grid.



\section*{Acknowledgments}

The authors acknowledge support from the SNCF grant No. PP00P2\_133552.

\bibliography{Electronic_transport_across_realistic_grain_boundaries_in_graphene}

\providecommand*\mcitethebibliography{\thebibliography}
\csname @ifundefined\endcsname{endmcitethebibliography}
  {\let\endmcitethebibliography\endthebibliography}{}
\begin{mcitethebibliography}{38}
\providecommand*\natexlab[1]{#1}
\providecommand*\mciteSetBstSublistMode[1]{}
\providecommand*\mciteSetBstMaxWidthForm[2]{}
\providecommand*\mciteBstWouldAddEndPuncttrue
  {\def\EndOfBibitem{\unskip.}}
\providecommand*\mciteBstWouldAddEndPunctfalse
  {\let\EndOfBibitem\relax}
\providecommand*\mciteSetBstMidEndSepPunct[3]{}
\providecommand*\mciteSetBstSublistLabelBeginEnd[3]{}
\providecommand*\EndOfBibitem{}
\mciteSetBstSublistMode{f}
\mciteSetBstMaxWidthForm{subitem}{(\alph{mcitesubitemcount})}
\mciteSetBstSublistLabelBeginEnd
  {\mcitemaxwidthsubitemform\space}
  {\relax}
  {\relax}

\bibitem[Gargiulo and Yazyev(2013)Gargiulo, and
  Yazyev]{gargiulo_topological_2013}
Gargiulo,~F.; Yazyev,~O.~V. Topological Aspects of Charge-Carrier Transmission
  across Grain Boundaries in Graphene. \emph{Nano Letters} \textbf{2013},
  \relax
\mciteBstWouldAddEndPunctfalse
\mciteSetBstMidEndSepPunct{\mcitedefaultmidpunct}
{}{\mcitedefaultseppunct}\relax
\EndOfBibitem
\bibitem[Novoselov et~al.(2004)Novoselov, Geim, Morozov, Jiang, Zhang, Dubonos,
  Grigorieva, and Firsov]{novoselov_electric_2004}
Novoselov,~K.~S.; Geim,~A.~K.; Morozov,~S.~V.; Jiang,~D.; Zhang,~Y.;
  Dubonos,~S.~V.; Grigorieva,~I.~V.; Firsov,~A.~A. Electric Field Effect in
  Atomically Thin Carbon Films. \emph{Science} \textbf{2004}, \emph{306},
  666--669\relax
\mciteBstWouldAddEndPuncttrue
\mciteSetBstMidEndSepPunct{\mcitedefaultmidpunct}
{\mcitedefaultendpunct}{\mcitedefaultseppunct}\relax
\EndOfBibitem
\bibitem[Geim and Novoselov(2007)Geim, and Novoselov]{geim_rise_2007}
Geim,~A.~K.; Novoselov,~K.~S. The rise of graphene. \emph{Nature Materials}
  \textbf{2007}, \emph{6}, 183--191\relax
\mciteBstWouldAddEndPuncttrue
\mciteSetBstMidEndSepPunct{\mcitedefaultmidpunct}
{\mcitedefaultendpunct}{\mcitedefaultseppunct}\relax
\EndOfBibitem
\bibitem[Castro~Neto et~al.(2009)Castro~Neto, Guinea, Peres, Novoselov, and
  Geim]{castro_neto_electronic_2009}
Castro~Neto,~A.~H.; Guinea,~F.; Peres,~N. M.~R.; Novoselov,~K.~S.; Geim,~A.~K.
  The electronic properties of graphene. \emph{Reviews of Modern Physics}
  \textbf{2009}, \emph{81}, 109--162\relax
\mciteBstWouldAddEndPuncttrue
\mciteSetBstMidEndSepPunct{\mcitedefaultmidpunct}
{\mcitedefaultendpunct}{\mcitedefaultseppunct}\relax
\EndOfBibitem
\bibitem[Coraux et~al.(2008)Coraux, N`Diaye, Busse, and
  Michely]{coraux_structural_2008}
Coraux,~J.; N`Diaye,~A.~T.; Busse,~C.; Michely,~T. Structural Coherency of
  Graphene on Ir(111). \emph{Nano Letters} \textbf{2008}, \emph{8},
  565--570\relax
\mciteBstWouldAddEndPuncttrue
\mciteSetBstMidEndSepPunct{\mcitedefaultmidpunct}
{\mcitedefaultendpunct}{\mcitedefaultseppunct}\relax
\EndOfBibitem
\bibitem[Yazyev and Louie(2010)Yazyev, and Louie]{yazyev_topological_2010}
Yazyev,~O.~V.; Louie,~S.~G. Topological defects in graphene: Dislocations and
  grain boundaries. \emph{Physical Review B} \textbf{2010}, \emph{81},
  195420\relax
\mciteBstWouldAddEndPuncttrue
\mciteSetBstMidEndSepPunct{\mcitedefaultmidpunct}
{\mcitedefaultendpunct}{\mcitedefaultseppunct}\relax
\EndOfBibitem
\bibitem[An et~al.(2011)An, Voelkl, Suk, Li, Magnuson, Fu, Tiemeijer, Bischoff,
  Freitag, Popova, and Ruoff]{an_domain_2011}
An,~J.; Voelkl,~E.; Suk,~J.~W.; Li,~X.; Magnuson,~C.~W.; Fu,~L.; Tiemeijer,~P.;
  Bischoff,~M.; Freitag,~B.; Popova,~E.; Ruoff,~R.~S. Domain (Grain) Boundaries
  and Evidence of {“Twinlike”} Structures in Chemically Vapor Deposited
  Grown Graphene. \emph{ACS Nano} \textbf{2011}, \emph{5}, 2433--2439\relax
\mciteBstWouldAddEndPuncttrue
\mciteSetBstMidEndSepPunct{\mcitedefaultmidpunct}
{\mcitedefaultendpunct}{\mcitedefaultseppunct}\relax
\EndOfBibitem
\bibitem[Huang et~al.(2011)Huang, Ruiz-Vargas, van~der Zande, Whitney,
  Levendorf, Kevek, Garg, Alden, Hustedt, Zhu, Park, McEuen, and
  Muller]{huang_grains_2011}
Huang,~P.~Y.; Ruiz-Vargas,~C.~S.; van~der Zande,~A.~M.; Whitney,~W.~S.;
  Levendorf,~M.~P.; Kevek,~J.~W.; Garg,~S.; Alden,~J.~S.; Hustedt,~C.~J.;
  Zhu,~Y.; Park,~J.; McEuen,~P.~L.; Muller,~D.~A. Grains and grain boundaries
  in single-layer graphene atomic patchwork quilts. \emph{Nature}
  \textbf{2011}, \emph{469}, 389--392\relax
\mciteBstWouldAddEndPuncttrue
\mciteSetBstMidEndSepPunct{\mcitedefaultmidpunct}
{\mcitedefaultendpunct}{\mcitedefaultseppunct}\relax
\EndOfBibitem
\bibitem[Kim et~al.(2011)Kim, Lee, Regan, Kisielowski, Crommie, and
  Zettl]{kim_grain_2011}
Kim,~K.; Lee,~Z.; Regan,~W.; Kisielowski,~C.; Crommie,~M.~F.; Zettl,~A. Grain
  Boundary Mapping in Polycrystalline Graphene. \emph{ACS Nano} \textbf{2011},
  \emph{5}, 2142--2146\relax
\mciteBstWouldAddEndPuncttrue
\mciteSetBstMidEndSepPunct{\mcitedefaultmidpunct}
{\mcitedefaultendpunct}{\mcitedefaultseppunct}\relax
\EndOfBibitem
\bibitem[Tapaszt\'{o} et~al.(2012)Tapaszt\'{o}, Nemes-Incze, Dobrik, Jae~Yoo,
  Hwang, and Bir\'{o}]{tapaszto_mapping_2012-1}
Tapaszt\'{o},~L.; Nemes-Incze,~P.; Dobrik,~G.; Jae~Yoo,~K.; Hwang,~C.;
  Bir\'{o},~L.~P. Mapping the electronic properties of individual graphene
  grain boundaries. \emph{Applied Physics Letters} \textbf{2012}, \emph{100},
  053114--053114--4\relax
\mciteBstWouldAddEndPuncttrue
\mciteSetBstMidEndSepPunct{\mcitedefaultmidpunct}
{\mcitedefaultendpunct}{\mcitedefaultseppunct}\relax
\EndOfBibitem
\bibitem[Rasool et~al.(2013)Rasool, Ophus, Klug, Zettl, and
  Gimzewski]{rasool_measurement_2013}
Rasool,~H.~I.; Ophus,~C.; Klug,~W.~S.; Zettl,~A.; Gimzewski,~J.~K. Measurement
  of the intrinsic strength of crystalline and polycrystalline graphene.
  \emph{Nature Communications} \textbf{2013}, \emph{4}\relax
\mciteBstWouldAddEndPuncttrue
\mciteSetBstMidEndSepPunct{\mcitedefaultmidpunct}
{\mcitedefaultendpunct}{\mcitedefaultseppunct}\relax
\EndOfBibitem
\bibitem[Lee et~al.(2013)Lee, Cooper, An, Lee, Zande, Petrone, Hammerberg, Lee,
  Crawford, Oliver, Kysar, and Hone]{lee_high-strength_2013}
Lee,~G.-H.; Cooper,~R.~C.; An,~S.~J.; Lee,~S.; Zande,~A. v.~d.; Petrone,~N.;
  Hammerberg,~A.~G.; Lee,~C.; Crawford,~B.; Oliver,~W.; Kysar,~J.~W.; Hone,~J.
  High-Strength Chemical-{Vapor–Deposited} Graphene and Grain Boundaries.
  \emph{Science} \textbf{2013}, \emph{340}, 1073--1076\relax
\mciteBstWouldAddEndPuncttrue
\mciteSetBstMidEndSepPunct{\mcitedefaultmidpunct}
{\mcitedefaultendpunct}{\mcitedefaultseppunct}\relax
\EndOfBibitem
\bibitem[Das~Sarma et~al.(2011)Das~Sarma, Adam, Hwang, and
  Rossi]{das_sarma_electronic_2011}
Das~Sarma,~S.; Adam,~S.; Hwang,~E.~H.; Rossi,~E. Electronic transport in
  two-dimensional graphene. \emph{Reviews of Modern Physics} \textbf{2011},
  \emph{83}, 407--470\relax
\mciteBstWouldAddEndPuncttrue
\mciteSetBstMidEndSepPunct{\mcitedefaultmidpunct}
{\mcitedefaultendpunct}{\mcitedefaultseppunct}\relax
\EndOfBibitem
\bibitem[Peres(2010)]{peres_colloquium:_2010}
Peres,~N. M.~R. Colloquium: The transport properties of graphene: An
  introduction. \emph{Reviews of Modern Physics} \textbf{2010}, \emph{82},
  2673--2700\relax
\mciteBstWouldAddEndPuncttrue
\mciteSetBstMidEndSepPunct{\mcitedefaultmidpunct}
{\mcitedefaultendpunct}{\mcitedefaultseppunct}\relax
\EndOfBibitem
\bibitem[Radchenko et~al.(2013)Radchenko, Shylau, Zozoulenko, and
  Ferreira]{radchenko_effect_2013}
Radchenko,~T.~M.; Shylau,~A.~A.; Zozoulenko,~I.~V.; Ferreira,~A. Effect of
  charged line defects on conductivity in graphene: Numerical Kubo and
  analytical Boltzmann approaches. \emph{Physical Review B} \textbf{2013},
  \emph{87}, 195448\relax
\mciteBstWouldAddEndPuncttrue
\mciteSetBstMidEndSepPunct{\mcitedefaultmidpunct}
{\mcitedefaultendpunct}{\mcitedefaultseppunct}\relax
\EndOfBibitem
\bibitem[Tsen et~al.(2012)Tsen, Brown, Levendorf, Ghahari, Huang, Havener,
  Ruiz-Vargas, Muller, Kim, and Park]{tsen_tailoring_2012}
Tsen,~A.~W.; Brown,~L.; Levendorf,~M.~P.; Ghahari,~F.; Huang,~P.~Y.;
  Havener,~R.~W.; Ruiz-Vargas,~C.~S.; Muller,~D.~A.; Kim,~P.; Park,~J.
  Tailoring Electrical Transport Across Grain Boundaries in Polycrystalline
  Graphene. \emph{Science} \textbf{2012}, \emph{336}, 1143--1146\relax
\mciteBstWouldAddEndPuncttrue
\mciteSetBstMidEndSepPunct{\mcitedefaultmidpunct}
{\mcitedefaultendpunct}{\mcitedefaultseppunct}\relax
\EndOfBibitem
\bibitem[Vancs\'{o} et~al.()Vancs\'{o}, M\'{a}rk, Lambin, Mayer, Kim, Hwang,
  and Bir\'{o}]{vancso_electronic_????}
Vancs\'{o},~P.; M\'{a}rk,~G.~I.; Lambin,~P.; Mayer,~A.; Kim,~Y.-S.; Hwang,~C.;
  Bir\'{o},~L.~P. Electronic transport through ordered and disordered graphene
  grain boundaries. \emph{Carbon} \relax
\mciteBstWouldAddEndPunctfalse
\mciteSetBstMidEndSepPunct{\mcitedefaultmidpunct}
{}{\mcitedefaultseppunct}\relax
\EndOfBibitem
\bibitem[Yu et~al.(2011)Yu, Jauregui, Wu, Colby, Tian, Su, Cao, Liu, Pandey,
  Wei, Chung, Peng, Guisinger, Stach, Bao, Pei, and Chen]{yu_control_2011}
Yu,~Q. et~al.  Control and characterization of individual grains and grain
  boundaries in graphene grown by chemical vapour deposition. \emph{Nature
  Materials} \textbf{2011}, \emph{10}, 443--449\relax
\mciteBstWouldAddEndPuncttrue
\mciteSetBstMidEndSepPunct{\mcitedefaultmidpunct}
{\mcitedefaultendpunct}{\mcitedefaultseppunct}\relax
\EndOfBibitem
\bibitem[Banhart et~al.(2011)Banhart, Kotakoski, and
  Krasheninnikov]{banhart_structural_2011}
Banhart,~F.; Kotakoski,~J.; Krasheninnikov,~A.~V. Structural Defects in
  Graphene. \emph{ACS Nano} \textbf{2011}, \emph{5}, 26--41\relax
\mciteBstWouldAddEndPuncttrue
\mciteSetBstMidEndSepPunct{\mcitedefaultmidpunct}
{\mcitedefaultendpunct}{\mcitedefaultseppunct}\relax
\EndOfBibitem
\bibitem[{READ} and {SHOCKLEY}(1950){READ}, and
  {SHOCKLEY}]{read_dislocation_1950}
{READ},~W.; {SHOCKLEY},~W. {DISLOCATION} {MODELS} {OF} {CRYSTAL} {GRAIN}
  {BOUNDARIES}. \emph{{PHYSICAL} {REVIEW}} \textbf{1950}, \emph{78},
  275--289\relax
\mciteBstWouldAddEndPuncttrue
\mciteSetBstMidEndSepPunct{\mcitedefaultmidpunct}
{\mcitedefaultendpunct}{\mcitedefaultseppunct}\relax
\EndOfBibitem
\bibitem[Yazyev and Louie(2010)Yazyev, and Louie]{yazyev_electronic_2010}
Yazyev,~O.~V.; Louie,~S.~G. Electronic transport in polycrystalline graphene.
  \emph{Nature Materials} \textbf{2010}, \emph{9}, 806--809\relax
\mciteBstWouldAddEndPuncttrue
\mciteSetBstMidEndSepPunct{\mcitedefaultmidpunct}
{\mcitedefaultendpunct}{\mcitedefaultseppunct}\relax
\EndOfBibitem
\bibitem[Van~Tuan et~al.(2013)Van~Tuan, Kotakoski, Louvet, Ortmann, Meyer, and
  Roche]{van_tuan_scaling_2013}
Van~Tuan,~D.; Kotakoski,~J.; Louvet,~T.; Ortmann,~F.; Meyer,~J.~C.; Roche,~S.
  Scaling Properties of Charge Transport in Polycrystalline Graphene.
  \emph{Nano Letters} \textbf{2013}, \emph{13}, 1730--1735\relax
\mciteBstWouldAddEndPuncttrue
\mciteSetBstMidEndSepPunct{\mcitedefaultmidpunct}
{\mcitedefaultendpunct}{\mcitedefaultseppunct}\relax
\EndOfBibitem
\bibitem[Wooten et~al.(1985)Wooten, Winer, and Weaire]{wooten_computer_1985}
Wooten,~F.; Winer,~K.; Weaire,~D. Computer Generation of Structural Models of
  Amorphous Si and Ge. \emph{Physical Review Letters} \textbf{1985}, \emph{54},
  1392--1395\relax
\mciteBstWouldAddEndPuncttrue
\mciteSetBstMidEndSepPunct{\mcitedefaultmidpunct}
{\mcitedefaultendpunct}{\mcitedefaultseppunct}\relax
\EndOfBibitem
\bibitem[Los and Fasolino(2003)Los, and Fasolino]{los_intrinsic_2003}
Los,~J.~H.; Fasolino,~A. Intrinsic long-range bond-order potential for carbon:
  Performance in Monte Carlo simulations of graphitization. \emph{Physical
  Review B} \textbf{2003}, \emph{68}, 024107\relax
\mciteBstWouldAddEndPuncttrue
\mciteSetBstMidEndSepPunct{\mcitedefaultmidpunct}
{\mcitedefaultendpunct}{\mcitedefaultseppunct}\relax
\EndOfBibitem
\bibitem[Li et~al.(2005)Li, Reich, and Robertson]{li_defect_2005}
Li,~L.; Reich,~S.; Robertson,~J. Defect energies of graphite:
  Density-functional calculations. \emph{Physical Review B} \textbf{2005},
  \emph{72}, 184109\relax
\mciteBstWouldAddEndPuncttrue
\mciteSetBstMidEndSepPunct{\mcitedefaultmidpunct}
{\mcitedefaultendpunct}{\mcitedefaultseppunct}\relax
\EndOfBibitem
\bibitem[Metropolis et~al.(1953)Metropolis, Rosenbluth, Rosenbluth, Teller, and
  Teller]{metropolis_equation_1953}
Metropolis,~N.; Rosenbluth,~A.~W.; Rosenbluth,~M.~N.; Teller,~A.~H.; Teller,~E.
  Equation of State Calculations by Fast Computing Machines. \emph{The Journal
  of Chemical Physics} \textbf{1953}, \emph{21}, 1087\relax
\mciteBstWouldAddEndPuncttrue
\mciteSetBstMidEndSepPunct{\mcitedefaultmidpunct}
{\mcitedefaultendpunct}{\mcitedefaultseppunct}\relax
\EndOfBibitem
\bibitem[Zakharchenko et~al.(2011)Zakharchenko, Fasolino, Los, and
  Katsnelson]{zakharchenko_melting_2011}
Zakharchenko,~K.~V.; Fasolino,~A.; Los,~J.~H.; Katsnelson,~M.~I. Melting of
  graphene: from two to one dimension. \emph{Journal of Physics: Condensed
  Matter} \textbf{2011}, \emph{23}, 202202\relax
\mciteBstWouldAddEndPuncttrue
\mciteSetBstMidEndSepPunct{\mcitedefaultmidpunct}
{\mcitedefaultendpunct}{\mcitedefaultseppunct}\relax
\EndOfBibitem
\bibitem[Lehtinen et~al.(2013)Lehtinen, Kurasch, Krasheninnikov, and
  Kaiser]{lehtinen_atomic_2013}
Lehtinen,~O.; Kurasch,~S.; Krasheninnikov,~A.~V.; Kaiser,~U. Atomic scale study
  of the life cycle of a dislocation in graphene from birth to annihilation.
  \emph{Nature Communications} \textbf{2013}, \emph{4}\relax
\mciteBstWouldAddEndPuncttrue
\mciteSetBstMidEndSepPunct{\mcitedefaultmidpunct}
{\mcitedefaultendpunct}{\mcitedefaultseppunct}\relax
\EndOfBibitem
\bibitem[Kurasch et~al.(2012)Kurasch, Kotakoski, Lehtinen, Skákalová, Smet,
  Krill, Krasheninnikov, and Kaiser]{kurasch_atom-by-atom_2012}
Kurasch,~S.; Kotakoski,~J.; Lehtinen,~O.; Skákalová,~V.; Smet,~J.;
  Krill,~C.~E.; Krasheninnikov,~A.~V.; Kaiser,~U. Atom-by-Atom Observation of
  Grain Boundary Migration in Graphene. \emph{Nano Letters} \textbf{2012},
  \emph{12}, 3168--3173\relax
\mciteBstWouldAddEndPuncttrue
\mciteSetBstMidEndSepPunct{\mcitedefaultmidpunct}
{\mcitedefaultendpunct}{\mcitedefaultseppunct}\relax
\EndOfBibitem
\bibitem[Mesaros et~al.(2010)Mesaros, Papanikolaou, Flipse, Sadri, and
  Zaanen]{mesaros_electronic_2010}
Mesaros,~A.; Papanikolaou,~S.; Flipse,~C. F.~J.; Sadri,~D.; Zaanen,~J.
  Electronic states of graphene grain boundaries. \emph{Physical Review B}
  \textbf{2010}, \emph{82}, 205119\relax
\mciteBstWouldAddEndPuncttrue
\mciteSetBstMidEndSepPunct{\mcitedefaultmidpunct}
{\mcitedefaultendpunct}{\mcitedefaultseppunct}\relax
\EndOfBibitem
\bibitem[Datta(1997)]{datta_electronic_1997}
Datta,~S. \emph{Electronic Transport in Mesoscopic Systems}; Cambridge
  University Press, 1997\relax
\mciteBstWouldAddEndPuncttrue
\mciteSetBstMidEndSepPunct{\mcitedefaultmidpunct}
{\mcitedefaultendpunct}{\mcitedefaultseppunct}\relax
\EndOfBibitem
\bibitem[Tapaszt\'{o} et~al.(2012)Tapaszt\'{o}, Nemes-Incze, Dobrik, Jae~Yoo,
  Hwang, and Bir\'{p}]{tapaszto_mapping_2012}
Tapaszt\'{o},~L.; Nemes-Incze,~P.; Dobrik,~G.; Jae~Yoo,~K.; Hwang,~C.;
  Bir\'{p},~L.~P. Mapping the electronic properties of individual graphene
  grain boundaries. \emph{Applied Physics Letters} \textbf{2012}, \emph{100},
  053114--053114--4\relax
\mciteBstWouldAddEndPuncttrue
\mciteSetBstMidEndSepPunct{\mcitedefaultmidpunct}
{\mcitedefaultendpunct}{\mcitedefaultseppunct}\relax
\EndOfBibitem
\bibitem[Koepke et~al.(2013)Koepke, Wood, Estrada, Ong, He, Pop, and
  Lyding]{koepke_atomic-scale_2013}
Koepke,~J.~C.; Wood,~J.~D.; Estrada,~D.; Ong,~Z.-Y.; He,~K.~T.; Pop,~E.;
  Lyding,~J.~W. Atomic-Scale Evidence for Potential Barriers and Strong Carrier
  Scattering at Graphene Grain Boundaries: A Scanning Tunneling Microscopy
  Study. \emph{ACS Nano} \textbf{2013}, \emph{7}, 75--86\relax
\mciteBstWouldAddEndPuncttrue
\mciteSetBstMidEndSepPunct{\mcitedefaultmidpunct}
{\mcitedefaultendpunct}{\mcitedefaultseppunct}\relax
\EndOfBibitem
\bibitem[Ihnatsenka and Zozoulenko(2013)Ihnatsenka, and
  Zozoulenko]{ihnatsenka_electron_2013}
Ihnatsenka,~S.; Zozoulenko,~I.~V. Electron interaction, charging, and screening
  at grain boundaries in graphene. \emph{Physical Review B} \textbf{2013},
  \emph{88}, 085436\relax
\mciteBstWouldAddEndPuncttrue
\mciteSetBstMidEndSepPunct{\mcitedefaultmidpunct}
{\mcitedefaultendpunct}{\mcitedefaultseppunct}\relax
\EndOfBibitem
\bibitem[Clark et~al.(2013)Clark, Zhang, Vlassiouk, He, Feenstra, and
  Li]{clark_spatially_2013}
Clark,~K.~W.; Zhang,~X.-G.; Vlassiouk,~I.~V.; He,~G.; Feenstra,~R.~M.;
  Li,~A.-P. Spatially Resolved Mapping of Electrical Conductivity across
  Individual Domain (Grain) Boundaries in Graphene. \emph{ACS Nano}
  \textbf{2013}, \emph{7}, 7956--7966\relax
\mciteBstWouldAddEndPuncttrue
\mciteSetBstMidEndSepPunct{\mcitedefaultmidpunct}
{\mcitedefaultendpunct}{\mcitedefaultseppunct}\relax
\EndOfBibitem
\bibitem[_la()]{_lammps_????}
LAMMPS Molecular Dynamics Simulator, http://lammps.sandia.gov\relax
\mciteBstWouldAddEndPuncttrue
\mciteSetBstMidEndSepPunct{\mcitedefaultmidpunct}
{\mcitedefaultendpunct}{\mcitedefaultseppunct}\relax
\EndOfBibitem
\bibitem[Plimpton(1995)]{plimpton_fast_1995}
Plimpton,~S. Fast Parallel Algorithms for Short-Range Molecular Dynamics.
  \emph{Journal of Computational Physics} \textbf{1995}, \emph{117},
  1--19\relax
\mciteBstWouldAddEndPuncttrue
\mciteSetBstMidEndSepPunct{\mcitedefaultmidpunct}
{\mcitedefaultendpunct}{\mcitedefaultseppunct}\relax
\EndOfBibitem
\end{mcitethebibliography}

\end{document}